# STEPWISE REDEFINITION OF THE SI BASE UNITS

**L K Issaev, S A Kononogov, V V Khruschov**

*VNIIMS, 46, Ozernaya St, 119361 Moscow, Russia*

**Abstract**

The four SI base units are proposed to be redefined in two stages: first, the kilogram, mole and ampere should be defined, and then the kelvin. To realize the redefinition of a base unit of the SI in terms of fundamental physical constant (FPC), a principle of coincidence of their physical dimensions is put forward. Direct applying this principle will lead to the changing of the sets of base and derived units of the new SI. If we want to preserve the continuity of the division between base and derived units in the new and the current SI, the principle is to be generalized with the time dimension factor be included. The status of the mole as the base unit of measurement is considered in the current and new SI. It is proposed to redefine the kilogram using a fixed value of the Avogadro constant and then to redefine the kelvin, after the measurement accuracy of the Boltzmann constant has been increased and agreed with the values of other constants of molecular physics.

---

## 1. Introduction

Resolution 1, adopted by the 24[th] CGPM, "On the possible future revision of the International System of units, the SI" invites "*the CIPM to continue its work towards improved formulations for the definitions of the SI base units in terms of fundamental constants, having as far as possible a more easily understandable description for users in general, consistent with scientific rigour and clarity.*" The aim of this paper is to discuss some procedural changes of the redefinition of the four SI base units (the kilogram, ampere, kelvin and mole) and clarification of formulations for the definitions of these SI base units in terms of FPCs, which can help to solve the stated in the 24th CGPM problems.

As is known, the proposals for definition of the four units in terms of the fixed values of FPCs in the new SI are based on the same principle of fixing of the exact value of the relevant FPC, which was used in 1983 for the new definition of the meter. Like this it is suggested to fix with a zero uncertainty exact values of the four constants $h$, $e$, $k_B$, and $N_A$, and then, by using them, to redefine the kilogram, ampere, kelvin and mole [1-11]. The proposal seems quite possible and is attractive, it has also disadvantages. They are worth being addressed again and a modified approach is offered here concerning both the choice of FPCs for the redefinition of the units, and, possibly, a different division between the base and derived units in the new SI.

## 2. Some advantages and disadvantages of fixing of constants' values

Let us consider in detail an example of fixing the value of the velocity of light $c$. The fixing of the $c$ value can be viewed as a successful metrological procedure for the definition of the base measurement unit - the unit of length. However, the introduction of the absolute metrological invariant – the value of the speed of light - implies that in the current SI it is possible to consider the speed rather than the length as a base unit of measurement. The length, here, is a derived unit, which is defined in terms of the speed of light and the unit of time. This is rather a matter of terminology (definitions of 'explicit constant' and 'explicit unit' [6]) and relies on the accepted agreement. It is also important that while introducing the definition of 'light meter' in 1983, the conditions for determining frequency and the speed of light with high accuracy were achieved, that is, one order of magnitude higher than the accuracy of measurement of the length, that allowed to fix the value of $c$.

However, the four constants $h$, $e$, $k_B$ and $N_A$ proposed for the redefinition of the SI units have another status today, and metrological characteristics of these constants have their own state of art. For this reason, it seems now premature to address the issue of fixing of all the four constants at the same time. Besides that, if all the new definitions of the SI units based on fixed values of $h$, $e$, $k_B$, and $N_A$ are adopted simultaneously, it will raise the problem of consistency of not only these but also of other constants. For instance, the simultaneous fixing of $h$, and $N_A$ will lead to a rigid connection for the constant of the speed of light in vacuum $c$, the molar mass constant $M_u$, the relative atomic mass of the electron $A_r(e)$, the fine-structure constant $\alpha$ and the Rydberg constant $R_\infty$. In order such relation does not happen, the existing definitions of some constants need to be changed, for example, the definition of the molar mass constant $M_u$. Therefore, the evident disadvantage of the new definitions consists in an additional change made to the definitions of constants, namely, transferring of $M_u$ into the category of variable physical quantities ($M_u = M_{u0}(1+\kappa)$), as suggested in the new definitions of the kilogram and mole [12].

The relation of $h$ and $N_A$, which includes the above-mentioned FPCs having higher accuracy ($\sim 10^9$) compared with the accuracy of the values of $h$ and $N_A$, is well known [13, 14]: $R_\infty / A_r(e)\alpha^2 = M_u c/2hN_A$, $M_u$ is equal to $10^{-3}$ kg/mol as $M_u = M(^{12}C)/12$, $M(^{12}C)$ is the molar mass of carbon-12. The constants $c$ and $M_u$ are fixed, whereas the $A_r(e)$, $R_\infty$ and $\alpha$ are the constants determined experimentally. Therefore, the fixing of values of $h$ and $N_A$ simultaneously at the current definitions of the constants will result in the additional relation for the values of $A_r(e)$, $R_\infty$ and $\alpha$, which does not comply with the physical laws.

## 3. Principle of correlation between dimensions of base units and nature invariants

The question is how to choose such FPCs whose values can be fixed for new definitions of the SI units without tension with physical laws and an existing large amount of measurement data. At first glance, it seems that the physical constant, whose value is determined by means of measurements, cannot be fixed and its value ever has a measurement uncertainty. However, if this uncertainty is less than a certain limit which depends on the accuracy of measurement, with the use of a SI unit, then for a new definition of the unit it is possible to fix the value of the chosen FPC, without contradicting existing measurement data. This metrological agreement, of course, has an impact, to a certain extent, on the physical meaning of the used constant.

Therefore, it is desirably that the number of fixed constants would be minimal and the link between the used FPC and the base measurement unit be simpler. This is possible, for example, if the base measurement unit and the used FPC have identical physical dimension. That is, if $[D]_{BU}$ is the dimension of a base unit, $[D]_{NI}$ is the dimension of an nature invariant, then $[D]_{BU} = [D]_{NI}$ is desirable.

Once this principle is applied, the meter and ampere will not be attributed to the base units of the new SI. The base units of measurement will then include the unit of speed and the unit of electric charge. Of course, it depends upon the adopted agreement what units are to be considered the base units. If we want to leave the meter and ampere among the base units of the new SI, then the above requirement has to be generalized.

Analogous to using the speed of light for the definition of the meter, the generalized condition for using of FPCs can be formulated as follows. A base unit of measurement is defined in terms of an invariant of nature, with the same dimension or with a dimension differing from the initial one by power of time:

$$[D]_{BU} = [D]_{NI} \times [T]^d, \qquad (1)$$

where $[D]_{BU}$ is the dimension of the base unit, $[D]_{NI}$ is the dimension of the invariant of nature, and $d$ is a rational number. As said earlier, the inclusion in the condition (1) of the factor with the dimension of power of time is related to maintaining of continuity of division into base and derived units in the current and new SI (the existing definition of the unit of length using a fixed value of the speed of light and a new definition of the ampere using the fixed value of the elementary charge). Taking into account the best accuracy of measurement quantities with the dimension of time or frequency, the possible presence of an additional factor $[T]^d$ does not impact the accuracy of reproducing of the unit under consideration.

Thus, the decision which of the two ways to use, i.e., to alter the sets of base and derived units of measurement or define a base unit in terms of an invariant of nature with the help of a time factor will depend on an adopted agreement. Let us remind of a fact that a set of base physical quantities consists of independent quantities, whereas a set of the base units of measurement of these quantities permits their dependence upon each other.

It is well known that, for instance, the definition of the mole in the existing SI explicitly depends on the definition of the kilogram. Nevertheless, the mole is the base unit of the amount of substance that can be only be defined after the kilogram has been defined. In the new SI it is possible to have peer units of mass and amount of substance. Moreover, the unit of the amount of substance can obtain more priority in the system of independent base units. Below possible definitions of the units of mass and amount of substance and a relationship between them will be considered.

## 4. Possible procedure for choosing constants to be fixed

Let us consider a procedure for choosing some FPCs, which are supposed to be fixed. When choosing certain FPCs it is necessary to take account of their status, the measurement accuracy of their values, as well as to bear in mind the continuity of new and historically formed magnitudes of the units. For example, the justification for fixing the value of the speed of light in vacuum was experimentally confirmed various consequences of the special theory of relativity, and the metrological ground was a higher accuracy of measurement of time periods compared with an accuracy of measurement of length intervals. Which physical quantities, besides the speed of light, today meet these requirements? First, these are the charges and masses of the existing molecules, atoms and elementary particles. The coincidence of physical characteristics of these entities has been confirmed not only in recent times, but far back as well by numerous experimental data. This allows considering of the physical characteristics of the molecules, atoms and particles as nature invariants. Therefore, we can fix the absolute value of the electric charge of the electron and, for example, the mass of the carbon-12 atom. The fixing of the mass of the carbon-12 atom of is in fact a choice of a microscopic measurement standard of mass.

Different proposals are known to have been made for new definitions of the units of mass and amount of substance. They were the subjects of discussion at meetings of the international metrological organizations and were presented at the 23rd and 24th CGPM. Among various criteria for transition to new definitions of the SI units the key one is the achievement of the level of $10^{-8}$ for the relative standard uncertainty of the values of Planck and Avogadro constants, and the coincidence of their values determined by different methods within the limits of the recommended standard uncertainties. If this can be obtained within the next two or three years, the issue of the choice of a particular definition of, say, the unit of mass, will probably remain open. In the present paper we suggest that the choice of the definition of any base unit of measurement should be made on the basis of the stated above generalized condition of correspondence of dimensions of a nature invariant and a measurement unit.

As is known, the earlier decisions of the CIPM and the 23rd CGPM concerning the replacement of the International Prototype Kilogram (IPK) [2, 5] have been suggested to be softened by the Consultative Committee for Mass and Related Quantities (CCM), in its latest recommendation G1, adopted at a meeting in 2010 [15], where the conditions for the new definition of the kilogram were formulated as follows:

- *At least three independent experiments, including work both from watt balance and from International Avogadro Coordination projects, yield values of the relevant constants with relative standard uncertainties not larger than 5 parts in $10^{-8}$. At least one of these results should have a relative standard uncertainty not larger than 2 parts in $10^{-8}$.*

- *For each of the relevant constants, values provided by the different experiments be consistent at the 95 % level of confidence.*

- *Traceability of BIPM prototypes to the international prototype of the kilogram be confirmed.*

The limits of the G1 recommendations are the top limits, since, as is shown in paper [16], a further increase in the stated above relative standard uncertainties of the Planck and Avogadro constants would violate the existing practice of measurement of mass with high accuracy for the masses of class $E_1$.

When introducing new definitions of the SI units the choice of the optimal number of FPCs should be made carefully, so that no logical or experimental discrepancies occur, and the highest level of accuracy is achieved. For the three constants $N_A$, $h$ and the atomic mass of carbon-12 it is possible to fix consistently only two constants [17, 18]. For example, in papers [11, 19] it was proposed to fix the atomic mass of carbon-12 and the constant $N_A$ [14, 20]. In this case it is not necessary to fix the value of $h$, as is proposed for the definition of the electronic kilogram [21]. It would be better to include the Planck constant to that class of constants, whose values are found by adjusting the results of various experiments. Because the use of a watt balance and introduction of the electronic kilogram, based on the quantum units of resistance and voltage, require a test of the consistency of the currently recommended values of the Josephson and von Klitzing constants with the help of devices based on single-electron tunneling, i.e. in fact, establishing a quantum current standard. Only then will it be possible to close the so-called 'quantum triangle' and substantiate the consistent system of practical electric units [22]. This 'closure' would mean the interconsistency of the values of the Planck constant, the elementary charge and the relations between the Josephson and von Klitzing constants, and is necessary for adopting and realization of the proposed new definition of the ampere with the simultaneously fixed values of $h$ and $e$. This is an argument in favour of the point that the adoption of the fixed value of $h$ is premature.

Really, an independent measurement of $h$ using watt balance, after the closure of 'the quantum triangle' would be an additional test of the relation $K_J^2 R_K = 4/h$ which has not yet been strictly proved, and hence the possible existence of a more general relation cannot be excluded where the corrections $\varepsilon_J$ and $\varepsilon_K$ are taken into account:

$$K_J^2 R_K (1 - 2\varepsilon_J - \varepsilon_K) = 4/h. \qquad (2)$$

Here the corrections $\varepsilon_J$ and $\varepsilon_K$ take into account the possible deviations $K_J$ and $R_K$ from the values of $K_{J0}=2e/h$ and $R_{K0}=h/e^2$, respectively, i.e. $K_J=K_{J0}(1+\varepsilon_J)$, $R_K=R_{K0}(1+\varepsilon_K)$ [19]. For the case of fixing of $N_A$ there arises an accurate method of measurement of $h$, using the values of the electron mass, the Rydberg constant and the fine-structure constant, and then the watt-balance method would be an accurate method for determining the product $K_J^2 R_K$, which would contribute to the development of the theory of Josephson and Hall macroscopic quantum effects.

It is desirable to set a minimum number of fixed FPCs required for measuring of physical quantities with high precision and stability. Fixing $h$ and $e$ at the same time, that is proposed in the CIPM recommendations and in a number of references, possibly, needs for solving of several tasks simultaneously - to introduce a new definition of the unit of mass, and to legitimize the use of the definitions of the electrical units on the basis of the Hall and Josephson quantum effects. However, the use of the practical electrical units $\Omega_{90}$ and $V_{90}$ within the SI system is possible but only after development of a quantum standard for electric current. So, we suggest fixing of the value of the Planck constant after development of a quantum standard for electric current and testing of the Josephson and von Klitzing relations at a higher level of accuracy using the new definitions of the units of mass, amount of substance and electric current.

As for the Boltzmann constant, the introduction of its fixed value today has not been caused by any practical need. It must be noted that in the foreseeable future, the existing accuracy of measurement of thermodynamic temperature will not be significantly improved and for practical measurements it is sufficient. Additionally, the exact value of the Boltzmann constant must be obtained from several experiments and have consistency with the values of other constants of molecular physics [23, 24]. The reasons mentioned above lead to the conclusion that in the nearest future it is well-grounded only to introduce the new definitions for the unit of electric current through a fixed charge of the electron, the units of mass and amount of substance on the basis of a fixed value of the Avogadro constant $N_A^*$ and the mass of the carbon 12 atom (the unit marked by asterisk hereinafter refers to the unit defined by using the fixed value of $N_A$).

The invariant, whose value is to be fixed by agreement, is the Avogadro constant $N_A$. When fixing a specific numerical value for $N_A$, i.e., the so-called Avogadro number $\{N_A\}$, we choose a traceable link between macro- and micro-world, which does not contradict any physical law. In order to maintain continuity with the current SI, it is convenient to keep the definition of the molar mass of carbon-12. For that, it is necessary to choose such number of carbon atoms, the total mass of which is the nearest to the mass of the IPK, taking account of the recommended uncertainty. This procedure will provide both a microscopic standard of mass, based on the mass of a carbon atom and a macroscopic standard of mass, based on the two invariants, namely, the mass of a carbon atom and a fixed value of $\{N_A\}$, i.e. $\{ N_A^* \}$.

## 5. The redefinitions of the mole and kilogram with the fixed value of the Avogadro constant

Let us consider in detail the procedure for redefining of the units of mass and amount of substance on the basis of a fixed value of the Avogadro constant $N_A^*$. As was said above, $N_A^*$ can be either chosen independent of the definition of the unit of mass, or related to it. For historical reasons, in the current SI the units of amount of substance and mass are in relation with each other. In this paper it is proposed to maintain this link in the new SI, that is, $N_A^*$ and the fixed number of carbon atoms ($N_{kg}^*$), which defines the macroscopic unit of mass in the new SI – the kilogram* - must be linked by the relation $\{N_A^*\} = 0{,}012\{N_{kg}^*\}$. It also means that the value of the molar mass constant $M_u$ in the new units (the kilogram* and mole*) is not changed [19, 25].

So, let us consider as the principal definition the following definition of the unit of mass: *the kilogram*, the unit of mass, is the exact mass $\{N_A^*\}/0{,}012$ of free atoms of carbon 12 at rest and in the ground quantum state.*

The obvious advantage of the suggested definition for the unit of mass is that it introduces at the same time (if the $M_u$ value is not changed) a new definition for the unit of amount of substance, namely: *the mole*, the unit of amount of substance, contains $\{N_A^*\}$ entities of the given substance.* This definition conforms to the currently existing SI definition.

It should be noted that such definition of the unit of amount of substance becomes independent of the definition of the kilogram and can be the first among definitions for base units. In this case the new definition of the kilogram is based on the invariant of nature - the mass of atom of carbon 12. It would be natural if the new and current units of mass of the SI and CGS, i.e., the kilogram and gram, contain an integer number of atoms of carbon-12. In this case, the value $\{N_A^*\}$ should be divisible by 12 [19]. Now the main task is to choose a value of $\{N_A^*\}$ from a certain interval, so that the relative difference between the value of the existing kilogram $\{K\}$ and the value of the new kilogram $\{K^*\}$ is of the order of $10^{-8}$. In this case, it is possible to retain the current values of the molar mass of carbon $M(^{12}C)$ and the molar mass constant $M_u$, being equal to 12 g/mol and 1 g/mol, respectively [19, 25]. Thus, the new molar mass of carbon and the molar mass constant will be expressed in terms of the kilogram ($kg^*$) and the mole

($mol$ *) analogous to the old ones, namely:

$$M(^{12}C)^* = 12 \times 10^{-3} \ кг^*/моль^*, \quad M_u^{\ *} = 10^{-3} \ кг^*/моль^*. \qquad (3)$$

These new definitions of the kilogram, mole, molar mass and the molar mass constant are not only convenient but necessary, since it does not destroy the established practice of evaluating of the mass and amount of substance in chemistry [25].

How can a certain fixed value of Avogadro number $\{N_A^*\}$ be chosen? This can be done today using the $\{N_A\}$ values - obtained with the help of the International Avogadro Coordination project (IAC) in experiments with silicon spheres - which belong to the 1-standard-deviation uncertainties interval (1σ-interval) [14]:

$$\{N_A\} = (6.02214066 \ ч \ 6.02214102) \ Ч10^{23}. \qquad (4)$$

In addition, recently there have been published the NRC results [26], where the 1σ-interval for Avogadro number is the interval of $(6.02214102 \ ч \ 6.02214140) \ Ч10^{23}$, however, there is a discrepancy between the values of the atomic mass of silicon, which were used by the IAC and NRC collaborations.

As can be seen there are several options of $N_A^*$. So, it makes sense to take into account more arguments while choosing a certain fixed value of $N_A^*$. For example, a geometric condition could be used, with the mole of carbon 12 filling a virtual cube [27]. In this case, $N_A^* = 84\ 446\ 889^3$. Also, special features of the structure of the silicon crystal lattice could be considered when choosing $\{N_A^*\}$. Though, in the surface layer of a silicon sphere the structure is much disrupted and needs to be restored with a computer simulation, the loss of accuracy is significant. In this connection it could be of interest a proposal to realize the mole or one gram of carbon by means of an even number of graphene sheets. The proposal can be actually realized in the foreseeable future. In [28] the parameters were calculated for a graphene hexagonal prism containing a certain number of carbon atoms, divisible by 12, and it was shown that the number nearest to Avogadro number can be chosen as follows:

$$\{N_A\} = 12(m/2)^2(1 + 3(m/2)) \qquad (5)$$

where $m = 51\ 150\ 060$. It can be seen that when m = 51 150 058, $\{N_A\}$ falls into the allowable $1\sigma$ interval (4), which has been obtained upon the completion of the Avogadro project [14]. So, the value $\{N_A^*\}$ can be chosen as follows,

$$\{N_A^*\} = 602\ 214\ 087\ 869\ 325\ 727\ 188\ 096. \qquad (6)$$

Finally, we have almost a unique set of definitions of the five base units of the new SI based on the fixed values of the FPCs, provided that for the definition of the kilogram and mole the fixed value of Avogadro number (6) is used.

*The second is the unit of time; it is the duration of 9192631770 periods of the radiation corresponding to the transition between the two hyperfine levels of the ground state of the caesium 133 atom at the zero thermodynamic temperature.*

*The meter is the length of the path travelled by light in vacuum during a time interval of 1/299 792 458 of a second.*

*The kilogram is the unit of mass; it is exactly equal to a sum of masses 50 184 507 322 443 810 599 088Ч10³ of unbound atoms of carbon 12, at rest and in their ground state.*

*The mole is the amount of substance of a system which contains 602 214 087 869 325 727 188 096 elementary entities of the given substance, exactly.*

*The ampere is the unit of electric current, the absolute value of the charge of the electron equals 1,602176487Ч10⁻¹⁹ coulombs, exactly.*

After the testing verification of the above definitions of the kilogram, mole and ampere have been completed, these definitions of the SI units will maintain continuity with the existing definitions; for the kilogram and mole, in particular, there will not be any miscount in the existing metrological chains of transfer of the units of mass, amount of substance, as well as in the common practice of measuring of mass and amount of substance [15, 16, 17, 18, 19, 25].

## 6. Conclusions

Intensive investigation has been conducting lately of the possibility of replacing the existing definitions of the four base SI units: the kilogram, ampere, kelvin and mole, with the definitions based on the exact values of the Planck, Avogadro and Boltzmann constants, and the elementary charge. Another main issue, besides the problem of increasing of the measurement accuracy of the above-mentioned constants, which has not yet been solved with the recommended level of accuracy, is the choice of the optimal set of FPCs, whose values are to be fixed for the new definitions of the units.

The opinion is most common today is that the new definition of the kilogram will be based on the fixed value of the Planck constant $h$, which is the fundamental constant of quantum mechanics, as $c$ is the fundamental constant of theory of relativity. Then the redefinition of the ampere through fixing of the electron charge $e$, when the values of $2e/h$ and $h/e^2$ have exactly defined values, will give advantage for the metrology of electrical measurements and will transfer the practical electrical units $\Omega_{90}$ and $V_{90}$ into the category of the SI units. However, the conclusion in this paper is such that the issue of fixing of $h$ will be possible only after the quantum current standard has been created.

The values that have to be fixed first of all are the values of $N_A$, the atomic mass of carbon 12 and the electron charge $e$, as the absolute metrological constants. The presented above procedure of fixing of $N_A$ and the atomic mass of carbon 12 leads to an explicit and logically correct definitions of the kilogram and mole, which are in concordance with the current definitions of these two units and the existing practice of measuring of mass and amount of substance. The fixing of $e$, taking into account the generalized principle of correspondence of dimensions of the base unit and the invariant of nature, allows to define the base unit – the ampere.

Thus, in this paper there have been proposed criteria for choosing the optimum set of FPCs with the fixed values for the redefinition of SI units and also, there has been presented a particular set of minimum specific FPCs which meets these requirements. This set, apart from the constant of ground state hyperfine transition of the caesium atom and the speed of light contains the Avogadro constant, the atomic mass of carbon 12 and the absolute value of the electron charge. The redefinition of the kelvin and candela, which essentially connected with the properties of macroscopic bodies, cannot be made now using generalized principle and taking account of the criteria proposed in this paper, it is beyond the considered approach. Further work is demanded as for making more precise new definitions of these units, as for obtaining needed precise FPCs values by several independent measurement methods. Therefore, the redefining of the four basic SI units - kilogram, ampere, kelvin and mole – is proposed to realize in two stages. At first, to redefine the kilogram, ampere, and mole on the basis of the fixed values of the Avogadro constant, the molar mass constant and the electron charge. And during the following stage, to redefine the kelvin (and, possibly, candela), after the accuracy of the Boltzmann constant has been increased by several measurement methods and agreed with the values of other constants of molecular physics.

# References

[1]   Mills I M *et al* 2005 Redefinition of the kilogram: a decision whose time has come
      *Metrologia* **42** 71-80

[2]   Bureau International des Poinds et Mesures 2005 *CIPM Recommendation 1*
      http://www.bipm.org

[3]   Mills I M *et al*  2006 Redefinition of the kilogram, ampere, kelvin and mole: a proposed
      approach to implementing  CIPM Recommendation 1 (CI-2005) *Metrologia* **43** 227-46

[4]   Bureau International des Poinds et Mesures 2008 *The International System of Units* 8th edn
      http://www.bipm.org

[5]    Bureau International des Poinds et Mesures 2007 *CGPM Resolution 12*
      http://www.bipm.org

[6]   Mills I M *et al*  2011 Adapting the International System of Units to the twenty-first century
      *Phil. Trans. R. Soc. Lond. Ser. A*  **369** 3907-24

[7]   Kononogov S A and Melnikov V N  2005 The fundamental physical constants, the
      gravitational constant, and the see space experiment project *Meas. Tech.* **48** 521-36

[8]   Kononogov S A 2008 Metrology and Fundamental Physical Constants *Moscow*
      *Standartinform (in Russian)*

[9]    Proc. Int. Meeting 2010 Fifty years of efforts toward quantum SI units *St. Petersburg*

[10]   Frantsuz E T 2010 Fundamental physical constants in the new SI *Meas. Tech.* **53** 228-31

[11]   Kononogov S A and Khruschov V V 2006 Scope for replacing the prototype kilogram by
       an atomic standard of the mass unit *Meas. Tech.* **49** 953-56

[12]   Taylor B N  2009 Molar mass and related quantities in the New SI *Metrologia* **46** L16-19

[13]   Mohr P J, Taylor B N and Newell D B 2008 CODATA recommended values of the
       fundamental physical constants: 2006  *Rev. Mod. Phys.* **80** 633-730

[14]    Andreas B *et al* (IAC) 2011 Counting the atoms in a $^{28}$Si crystal for a new kilogram

definition *Metrologia* **48** S1-13

[15]    Bureau International des Poinds et Mesures 2010 CCM *Recommendation G1*

        http://www.bipm.org

[16]    Glдser M *et al* 2010 Redefinition of the kilogram and the impact on its future

        dissemination *Metrologia* **47** 419-28

[17]    Leonard B P 2010 Comments on recent proposals for redefining the mole and kilogram

        *Metrologia* **47** L5-8

[18]    Jeannin Y 2010 What is a mole?: old concepts and new (continued) *Chem. Int.* **32**(1) 8-11

[19]    Khruschov V V  2010 Possible definition of the unit of mass and fixed values of the

        fundamental physical constants  *Meas. Tech.* **53** 583-91; *ibid.* **54** 1103-10

[20]    Valkiers S *et al*  2011 SI primary standards for the calibration of ion-current ratios in the

        molar-mass measurement of natural Si single crystals *Metrologia* **48** S26-31

[21]    Steiner R, Williams E R and Newell D B 2005 Towards an electronic kilogram: an

        improved measurement of the Plank constant and electron mass *Metrologia* **42** 431-41

[22]    Wulf M and Zorin A B  2008 Error accounting in electron counting experiments

        *arXiv. – 0811.3927* 1-12

[23]    Kalinin M I and Kononogov S A  2010 Redefinition of the Unit of Thermodynamic

        Temperature in the SI System *High Temperat.* **48** 23-28

[24]    Bureau International des Poinds et Mesures 2010 *CCT Recommendation T2*

        http://www.bipm.org

[25]    Leonard B P 2012 Why the dalton should be redefined exactly in terms of the kilogram

        *Metrologia* **49** 487-91

[26]    Steele A G *et al*  2012  Reconciling Planck constant determinations via watt balance and

        enriched-silicon measurements at NRC Canada *Metrologia* **49** L8-10

[27]    Hill T P, Miller J and Censullo A C  2011 Towards a better definition of the kilogram

        *Metrologia* **48** 83-86

[28]     Fraundorf P  2012 A multiple of 12 for Avogadro *arXiv:1201.5537* 1-3